\documentclass[11pt,twoside]{article}

\usepackage{asp2004}
\usepackage{epsf}
\usepackage{psfig}
\usepackage{lscape}
\usepackage{graphicx}

\begin{document}
\title{SINS of Viscosity Damped Turbulence}   %%% Fill in title
\author{A. Lazarian }   %%% Fill in author names
\affil{Astronomy Department, University of Wisconsin-Madison}    %%% Fill in author affiliations

\begin{abstract} %%% Abstract to run on from here.
The problems with explaining the Small Ionized and Neutral Structures (SINS)
appealing to turbulence
stem from inefficiency of the Kolmogorov cascade in creating
 large fluctuations at sufficiently small scales. However, other
types of cascades are possible.
When magnetic turbulence in a fluid with viscosity that is much larger
than resistivity gets to a viscous damping scale, the turbulence does 
not vanish. Instead, it gets into a different new regime.
Viscosity-damped turbulence produces fluctuations on the small scales.
Magnetic fields sheared by turbulent motions by eddies not damped by
turbulence create small scale filaments that are confined by the
external plasma
pressure. This creates small scale density fluctuations. In addition,
extended current sheets create even stronger density gradients
that accompany field reversals in the plane
perpendicular to mean magnetic field. Those
 can be responsible for the SINS formation. This scenario is applicable to
partially ionized gas. 
More studies of reconnection in the viscosity dominated regime are necessary
to understand better the extend to which the magnetic reversals can compress
the gas.
\end{abstract}

%%% MAIN BODY OF TEXT GOES HERE. CONSULT "INSTRUCTIONS FOR AUTHORS USING
%%% LATEX2E MARKUP", SECTIONS 2.3-2.6 FOR HELP WITH EQUATIONS, FIGURES,
%%% AND TABLES.
\section{Structures below Viscous Damping Scale}

Turbulence can be viewed as a cascade of energy from a large 
injection energy scale
to dissipation at a smaller scale. The latter is being established by equating 
the rate of
turbulent energy transfer to the rate of energy damping arising, for instance, 
from viscosity. Naively, one does not  expect to see any turbulent phenomena 
below such a scale.

Such reasoning may not be true in the presence of magnetic
field, however. Consider magnetized fluid with 
viscosity $\nu$ much
larger than magnetic diffusivity $\eta$, which is the case of
a high magnetic
Prantl number $Pr$ fluid.
The partially ionized gas can serve as an example of such a fluid
up to the scales of ion-neutral decoupling (see a more rigorous 
treatment in Lazarian, Vishniac \& Cho 2004, henceforth LVC04).
Fully ionized plasma is a more controversial example. For
instance, It is well known that
for plasma the diffusivities are different along and perpendicular 
to magnetic field lines. Therefore, the plasma the 
Prandtl number is huge if we use the parallel diffusivity 
$\nu_{\|}\gg \nu_{\bot}$. A treatment of the fully ionized plasma
as a high Prandtl number medium is advocated in Schechochihin et al (2004,
henceforth SCTMM).
 
The turbulent cascade in a fluid with isotropic $\eta$ proceeds up
to a scale at which the cascading rate, which for the Kolmogorov
turbulence, i.e. $v_l\sim l^{1/3}$, 
is determined by the eddy turnover rate $v_l/l$ gets equal
to the damping rate $\eta/l^2$. Assuming that the energy is injected
at the scale $L$ and the injection velocity is $V_L$, the damping
scale $l_c$ is $L Re^{-3/4}$, where $Re$ is the Reylonds number $L V_L/\eta$.
  
However, it is evident that magnetic
fields at the scale $l_c$ 
at which hydrodynamic cascade would stop are still sheered by
eddies at the larger scales. This should result in creating magnetic
structures
at scales $\ll l_c$ (LVC04). Note, that magnetic field in this regime
is not a passive scalar, but important dynamically.

Cho, Lazarian \& Vishniac (2002) 
reported a new viscosity-damped regime of turbulence
using incompressible MHD simulations (see an example of
more recent simulations in Fig.~1). In this regime, unlike 
hydro turbulence, motions, indeed, do not stop at the viscosity damping scale,
but magnetic fluctuations protrude to smaller scales. Interestingly
enough, these magnetic fluctuations induce small amplitude
 velocity fluctuations at scales $\ll l_c$.
 
Cho\& Lazarian (2003)
confirmed these results with compressible simulations and 
speculated that these small scale magnetic
fluctuations can compress ambient gas to produce SINS (see also Fig. 1b). 
According to the model in LVC04, while the
 spectrum of averaged over volume magnetic fluctuations scales as $E(k)\sim
k^{-1}$, the pressure within intermittent magnetic structures increases 
with the decrease of the scale as $(\delta \hat{b})^2_k\sim k$, while 
the filling factor 
$\phi_k \sim k^{-1}$,
the latter being consistent with numerical simulations.
%\footnote{Both
%spectra of magnetic field and velocity in numerical simulations somewhat
%steeper than expected. This difference was shown in LVC04 to correspond to
%the expectations related to the effects of the limited dynamical
%range.}. 
The pressure of gas 
confining the magnetic filaments should increase accordingly, resulting in 
fluctuations of density increasing with the decrease of the scale. 
The fact that the emerging structures are filamentary allows occasional picks 
in the observed column densities, which could correspond to the Heiles (1987) 
model of SINS.    
\begin{figure}
  \includegraphics[width=3.6cm]{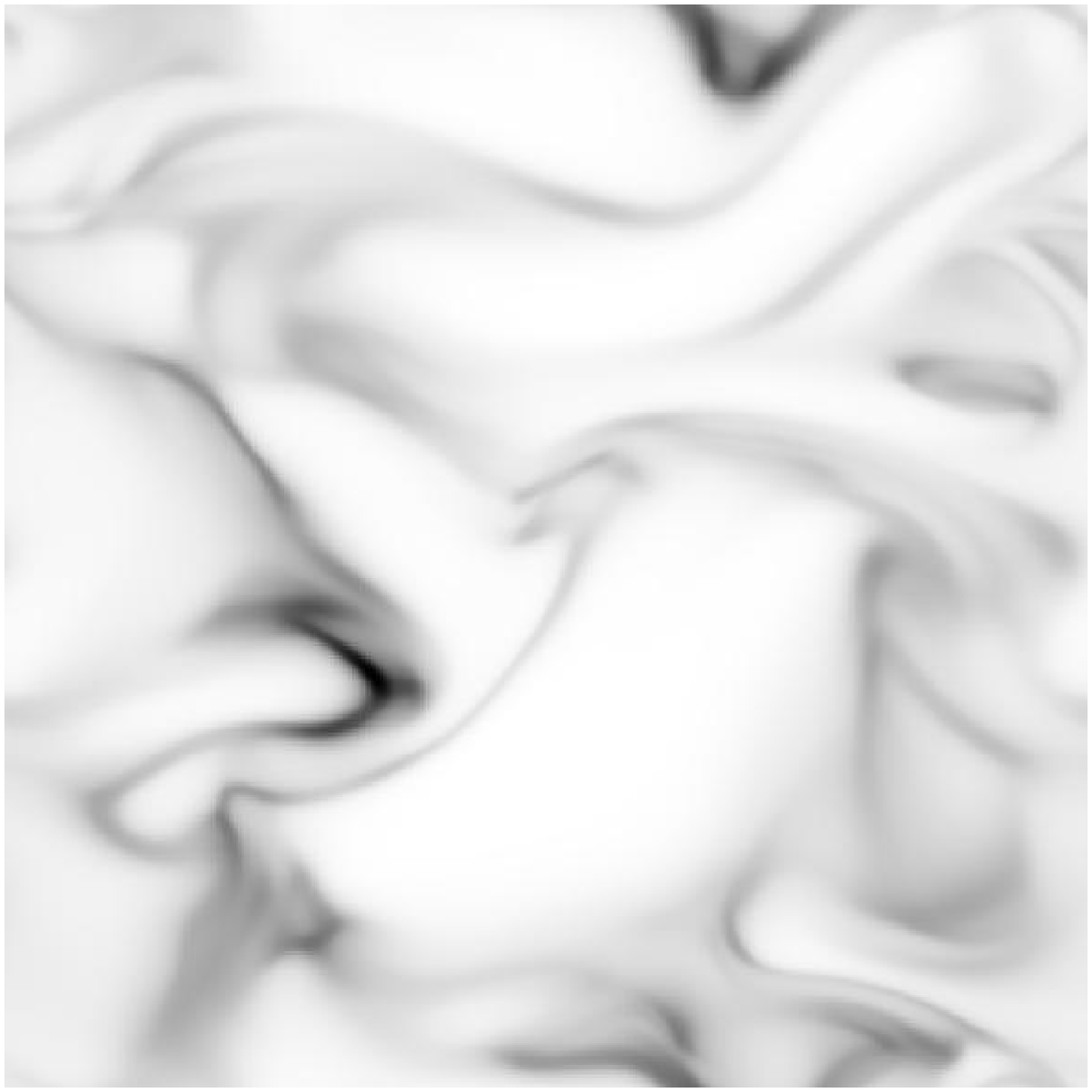}
  \hfill
  \includegraphics[width=0.67cm]{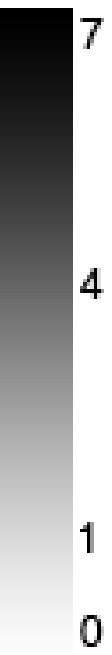}
 % \hfill
\hspace{0.07cm}\hfill
  \includegraphics[width=5.0cm]{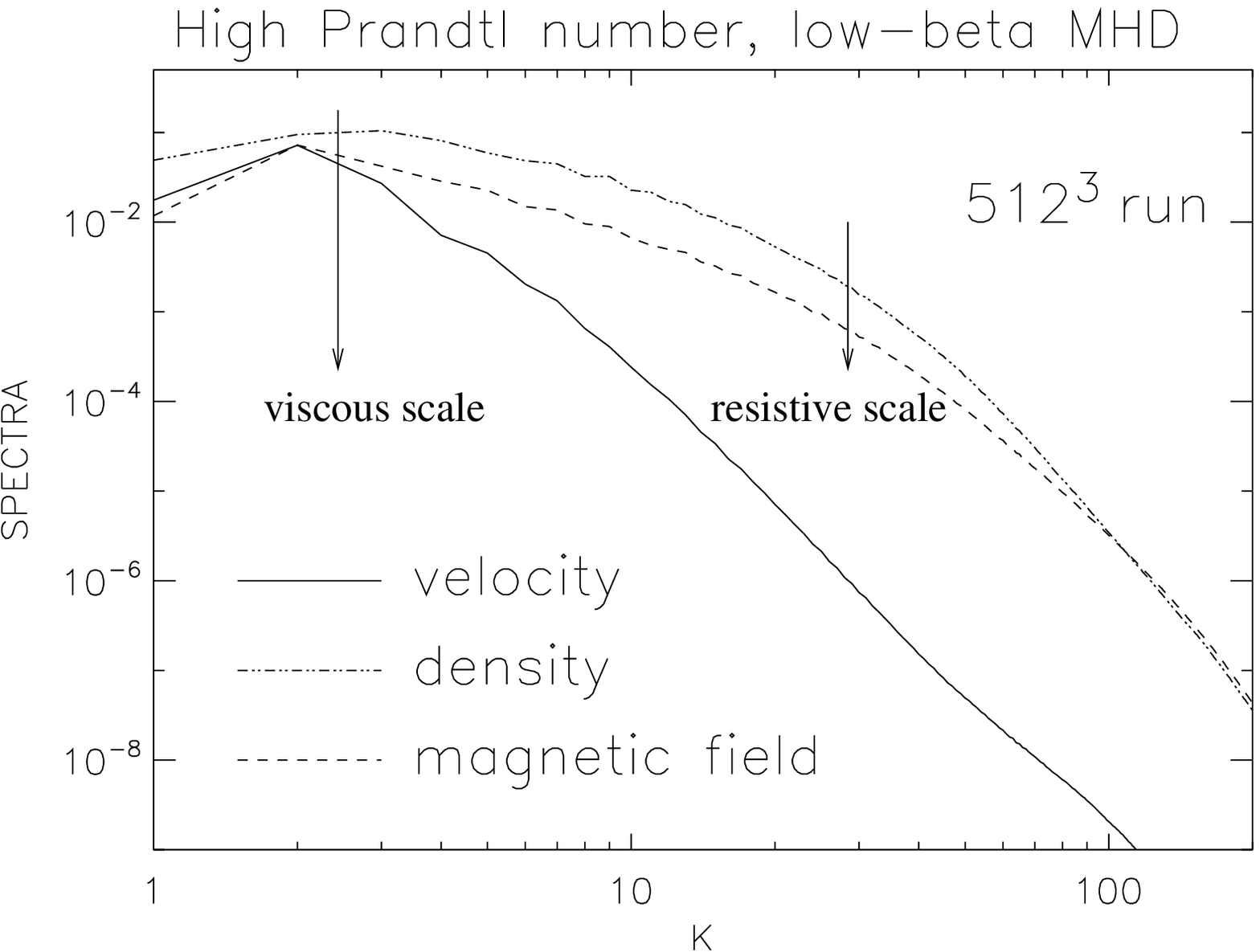}
\caption{ {\it Left}: Filaments of density created by magnetic
compression in the slice of data cube of the viscosity-damped regime of MHD turbulence. 
{\it Right}: Spectra of density and magnetic field are similar,
while velocity is damped. The resistive scale in this regime is not
$L/Rm$ but $L {Rm}^{-1/2}$.(from Beresnyak \& Lazarian, in preparation).}
\label{visc}
\end{figure}

In Fig.~1 we show the results of
our high resolution compressible MHD simulations that exhibit strong density 
fluctuations at the scales
below the one at which hydrodynamic turbulence would be damped. 
 The testing of the LVC04 model 
is important by itself, even without a possible connection to SINS. 
According to  that model, the viscosity-damped
regime is ubiquitous in turbulent partially ionized gas. Some of its 
discussed consequences, such as an intermittent resumption of the turbulence
the fluid of ions as magnetic fluctuations reach the ion-neutral decoupling
scale, are important for radio scintillations.

\section{Current Sheets in Viscosity-Damped Regime}

We note, that irrespectively of the density fluctuations arising 
from compressions by magnetic
fields, the viscosity-damped turbulence produces current sheets with length 
determined by
the size of the eddies at the viscous damping scale $l_c$.
If we assume that the thickness of the resulting current sheets
is  determined by the Sweet-Parker (henceforth SP)
reconnection condition (see Parker 1979)
 and therefore is $d\sim l_c Rm^{-1/2}$,
where $Rm$ is the Lunquist number for a scale $l_c$, i.e.
$\eta/V_A l_c$, where $V_A$ is the Alfven velocity.
In the SP reconnection the
magnetic pressure changes across the current sheet by $(\delta b)_c^2$,
which entails the corresponding variation of density $\delta n/n\sim
(\delta b_c)^2/P_{gas}$. Thus for favorable observing geometry
the current layer induces the maximal variation of observed column density
at the scale of $d$ which is $l_c (\delta b)_c^2$. This value is
a factor of $(l_c/d)^{5/3}\sim Rm^{5/6}$ larger than the column density of
an eddy at a scale 
 created
at the scale $d$, provided that the compression is due to
$\delta b$, which cascading proceeds according to 
the Kolmogorov law. 

Assume, for the sake of simplicity, that at the injection scale the injection
velocity is equal to $V_A$. This is the situation for which the 
Goldreich-Sridhar (1995) model of MHD turbulence has been formulated
originally. In some situations the discussion of a more general case
is essential (see Lazarian 2006), but this goes beyond the scope of this short 
communication.
In terms of the magnetic Prandtl number the amplification
of density perturbations at a scale $d$ is of the order 
$Pr^{5/6} Re^{5/18}$.
If, following Goldreich \& Sridhar (2006, henceforth GS06), we use for plasma 
$Pr\approx 3\times 10^8 (T/10^4 {\rm K})$, then an eddy at scale $d$
can provide a density contrast that is 
$\sim 10^{7} (T/10^4 {\rm K})^{5/6} Re^{5/18}$ times larger than the
 Kolmogorov prediction.

\begin{figure}
  \includegraphics[width=6.2 cm]{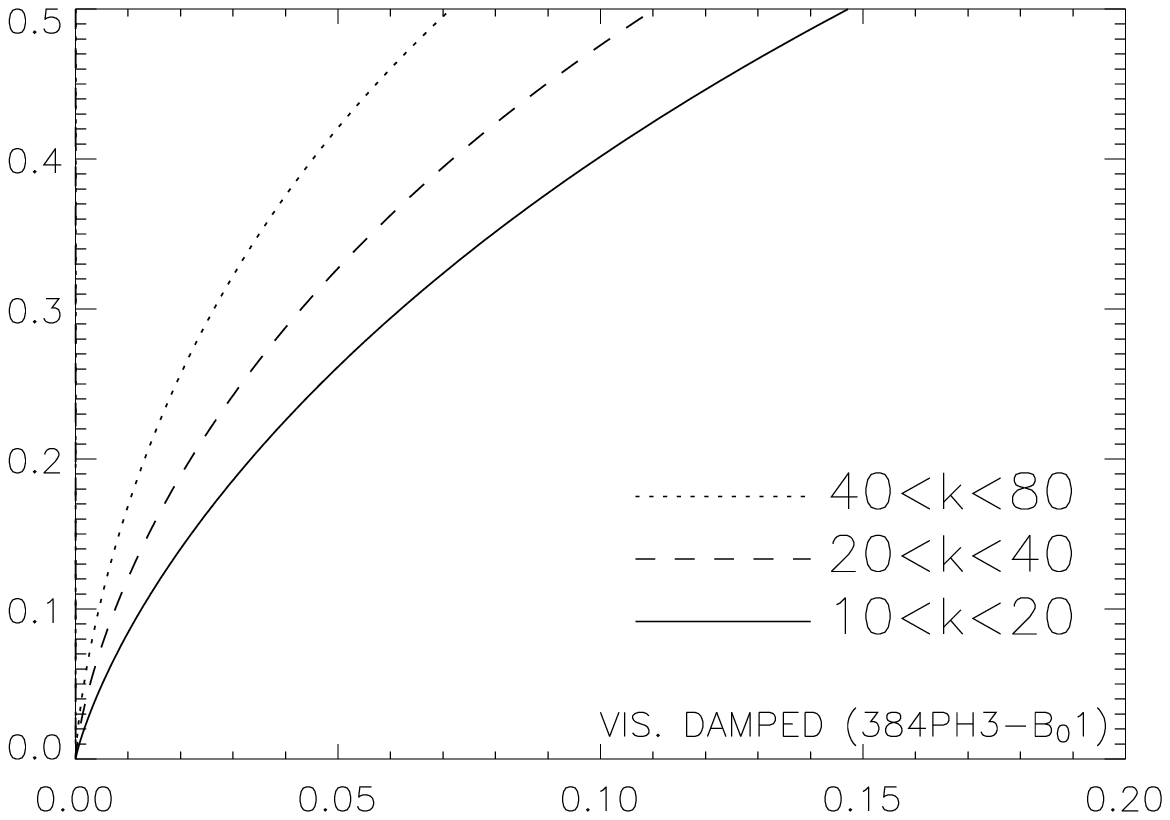} 
  \hfill
  \includegraphics[width=4.5cm]{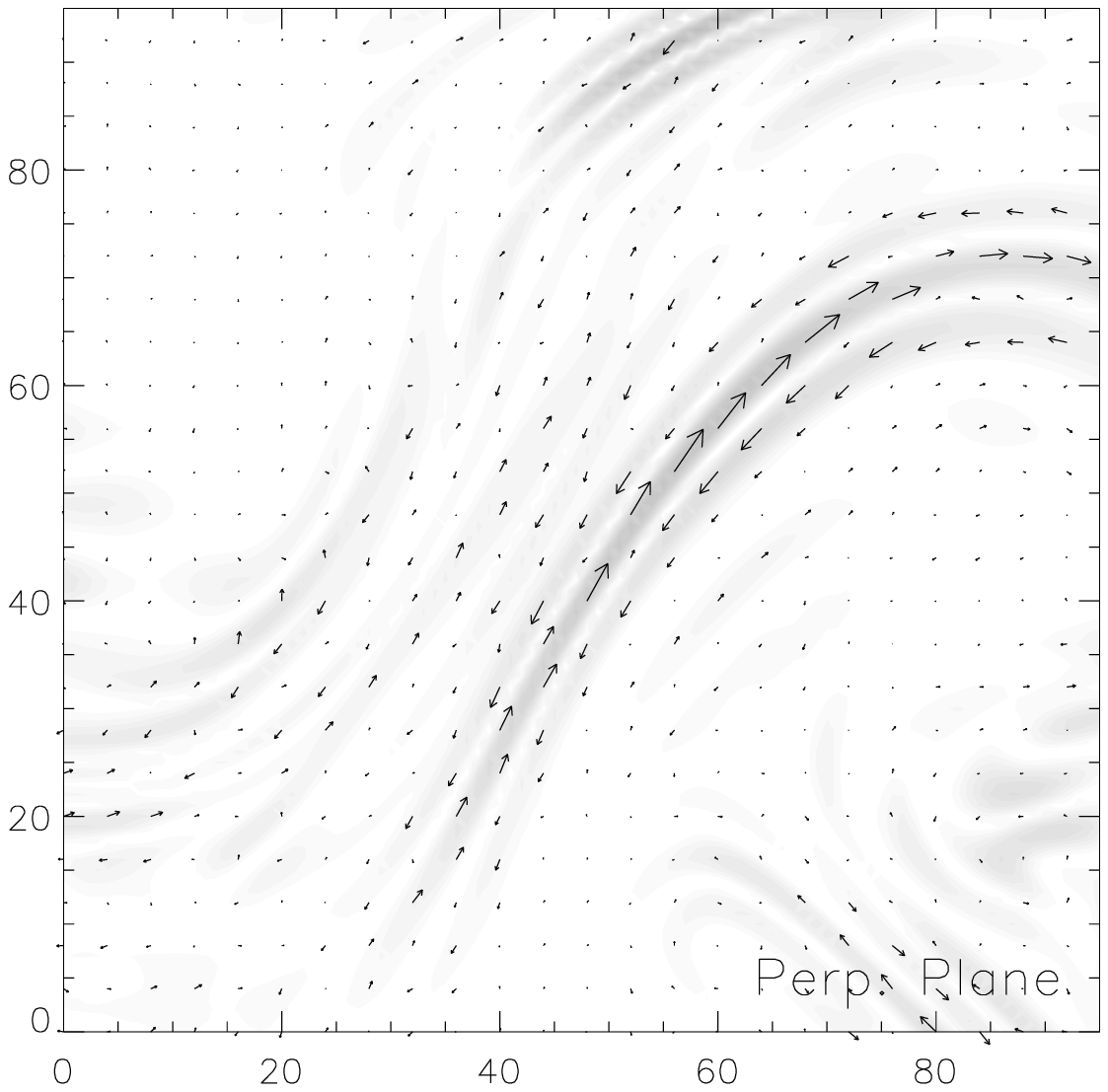}
\caption{ {\it Left}: Fractional volume (X-axis) versus fractional magnetic
energy in the volume (Y-axis) for viscosity-damped MHD turbulence: smaller
scales show higher concentration of magnetic energy (from LVC04)
{\it Right}: Strength of magnetic field in a plane perpendicular to the
mean magnetic field. Magnetic field reversals are clearly seen (from
Cho, Lazarian \& Vishniac 2003). In these
planes the structure of the magnetic field resembles the structure of
the folded fields in SCTMM, while the 3D structure
of the magnetic field is different in those two cases. Note, that the
reversals of magnetic fields reported for different media using
Faradey rotation technique may be due to observations 
with the line of sight perpendicular to the mean magnetic field.}
\label{visc}
\end{figure}

\section{Discussion}

Recently GS06 attempted to explain the extreme radio-wave scattering 
in the direction of the Galactic Center assuming plasma can be
described with high $Pr$ number and the folded
fields in the spirit of SCTMM are present along the line of sight.
They do not appeal directly to the SP reconnection process, but the
arguments about the current sheets there implicitly assume that
the SP-type reconnection. As the result, their estimates of the
variation of the column density can be obtained from ours assuming
that $L=l_c$ for the folded fields model.
Compared to the GS06 the model in the paper is applicable to the situations
when strong mean magnetic field is present. If attempting to explain the
extreme scattering events with viscosity-damped turbulence
 model, one could use the corresponding estimates
in GS06, modifying the column density fluctuation as discussed above.
In the highly viscous fluid the values of $L$ and $l_c$ do not differ
dramatically, anyhow.

GS06 and the model above when
 applied to plasmas share the same set of problems. 
As GS06 admitted, the SCTMM folded field structure could unwind rapidly
due to the low kinematic viscosity in directions perpendicular to the
magnetic field. In view of this, the formation of the SINS in partially
ionized gas (e.g. HI) could be seen as a safer bet.

Another potential difficulty shared by both models
is related to using SP model of current sheets
for both the folded fields and the viscosity damped turbulence. 
Usually, one would expect that the reconnection happens faster,
opening up the reconnection layers (see Shay et al. 2001 for
collisionless reconnection and Lazarian \& Vishniac 1999 for 
stochastic reconnection). However, the conditions for the field
wandering and the outflows within both the viscosity damped turbulence
at scales less than $l_c$ (see
LVC04)
and the folded field dynamo (see SLTMM) differ from the usual
formulation of the reconnection problem. Further research should
clarify what what is going on in such situations. Potentially,
compression in current sheets looks as an attractive solution of
the SINS phenomenon.

\acknowledgements %%% Text of acknowledgements runs on after this command.
AL acknowledges the NSF grant
AST-0307869 and the support from the Center for Magnetic
Self-Organization in Laboratory and Astrophysical Plasmas. 

%%% THE BIBLIOGRAPHY
%%%
%%% CONSULT SECTION 3 OF "INSTRUCTIONS FOR AUTHORS" FOR HOW TO USE NATBIB.
%%% AUTHORS ARE ENCOURAGED TO USE EITHER THE "THEBIBLIOGRAPY" ENVIRONMENT
%%% BY UNCOMMENTING (DELETING THE "%" SYMBOL) THE COMMANDS BELOW, OR BY
%%% USING THE BIBTEX ENVIRONMENT. TO FIND OUT WHICH IS APPLICABLE TO YOUR
%%% CONTRIBUTION, CONSULT THE VOLUME EDITORS FOR YOUR PROCEEDINGS.
%%%

\end{document}